\documentclass[10pt, conference, letterpaper]{IEEEtran}
\IEEEoverridecommandlockouts
\usepackage{cite}
\usepackage{amsmath,amssymb,amsfonts}
\usepackage{algorithmic}
\usepackage{graphicx}
\usepackage{textcomp}
\usepackage{xcolor}
\usepackage[ruled,linesnumbered]{algorithm2e}
\usepackage{subfigure}
\usepackage{booktabs}
\def\BibTeX{{\rm B\kern-.05em{\sc i\kern-.025em b}\kern-.08em
    T\kern-.1667em\lower.7ex\hbox{E}\kern-.125emX}}

\newcommand{\tabincell}[2]{\begin{tabular}{@{}#1@{}}#2\end{tabular}}

\begin{document}

\title{SegMobaTree: The Segmented Multilayer Online Balanced Tree for high-performance IPv6 Lookup in the Edge Network}




\author{\IEEEauthorblockN{Chunyang Zhang \IEEEauthorrefmark{1}\IEEEauthorrefmark{3}, Gaogang Xie \IEEEauthorrefmark{2}\IEEEauthorrefmark{3}}
\IEEEauthorblockA{
\IEEEauthorrefmark{1}Institute of Computing Technology, Chinese Academy of Sciences, China\\
\IEEEauthorrefmark{2}Computer Network Information Center, Chinese Academy of Sciences, China\\
\IEEEauthorrefmark{3}University of Chinese Academy of Sciences, China \\
zhangchunyang@ict.ac.cn, xie@cnic.cn}}

\maketitle


\begin{abstract}

With the development of IPv6 and edge computing, the edge network should support IPv6 lookup (the longest prefix matching, LPM) with high lookup speed, high update speed, and low memory cost. However, the trie-based algorithms, e.g., SAIL and Poptrie, mainly focus on the IPv4 ruleset but have disadvantages in the edge IPv6 ruleset with longer prefix length. The binary-based algorithm Hi-BST also has limited lookup speed with too many memory accesses. Therefore, we propose the SegMobaTree algorithm to achieve high-performance IPv6 lookup in the edge network. First, MobaTree is a multilayer online balanced tree to perform high-speed binary search among rules with different prefix lengths. Second, to avoid one large tree, we propose the dynamic programming method to split prefix lengths into a few suitable segments, which tradeoff between the number of segments and the scale of trees. Compared to SAIL, Poptrie, and Hi-BST, SegMobaTree achieves 1.5x, 1.7x, 1.6x lookup speed and 11.8x, 32.8x, 1.3x update speed with low memory cost.

\end{abstract}



\section{Introduction}

IP lookup is the core component for network devices, such as routers and switches, to support packet transmission in the network. For each packet, the device will extract the destination IP address and lookup a matching rule with the longest prefix length, then transmit the packet to the port according to this matching rule. Thus the IP lookup is a longest prefix matching (LPM) problem. 

Considering the development of IPv6 and edge computing, the IPv6 lookup in the edge network becomes the bottleneck, but it is hard to be improved. First, the longer prefix lengths of ruleset limit the lookup speed but the increasing traffic requires high-speed lookup as IPv4. Second, the complex and variant network require high-speed rule updates. Third, the memory space of device is limited but the scale of IPv6 ruleset is increasing quickly in recent years. Therefore, the IPv6 lookup algorithm in the edge network should perform high-speed lookup and update with low memory cost. 

Although lots of algorithms are proposed, none can achieve high lookup speed, high update speed, and low memory cost simultaneously. Some algorithms can not be applied in IPv6 lookup like Dir24-8 \cite{Dir-24-8} and DXR \cite{DXR}. Some trie-based algorithms like LC-Tries \cite{LCTries}, Tree-bitmap \cite{TreeBitmap}, SAIL \cite{Sail}, and Poptrie \cite{Poptrie} have disadvantages in IPv6 rulesets with longer prefix lengths. The binary-based algorithm Hi-BST \cite{Hi-BST} can perform high-speed update with low memory cost, but its lookup speed is limited beacuse the inefficient data structure produces too many memory accesses. Therefore, the existing algorithms are insufficient to support current IPv6 lookup in the edge network. 

To achieve high performance for handling IPv6 rules with long different prefix lengths, we propose the multilayer online balanced tree MobaTree. In each layer, it constructs a balanced tree to classify non-intersect rules. If some rules are contained by another rule, it further constructs a balanced tree for them in the next layer of this rule. With the multilayer structure, MobaTree consumes low memory cost and achieves high lookup and update speeds with $O(log_2(n))$ complexity.

To avoid one large tree and achieve higher performance, we propose the dynamic pogramming method to split rules with different prefix length into a few segments. Each segment will apply a hash table to classify rules with different reduced prefix into different small trees. Because the lookup process will traverse segments and then lookup in the matching trees, we use dynamic programming to find suitable segments, which tradeoff the memory accesses bewteen the number of segments and the scale of trees. 

The experimental results show that SegMobaTree has the highest lookup and update speeds with low memory cost simultaneously. Compared to SAIL, Poptrie, and Hi-BST, SegMobaTree achieves 1.5x, 1.7x, 1.6x lookup speed and 11.8x, 32.8x, 1.3x update speed with low memory cost as Hi-BST.


\section{Related Work}

Existing IPv6 lookup algorithms can be divided into three categories. The first category is the hardware-based algorithms that improve the lookup speed but consume additional hardware cost like TCAM \cite{2005Algorithms, Hybrid} and FPGA \cite{SRAM-based, Scalable}. The second category is the trie-based algorithms that only focus on high-speed IPv4 lookup. The third category is the binary-based algorithms with high update speed and low memory cost but limited lookup speed. Our proposed SegMobaTree falls into the third category. However, through the efficient multilayer online balanced structure and the suitable segments, it achieves high lookup speed, high update speed and low memory cost simultaneously.

\textbf{The trie-based algorithms:}
By lookup through the first bit until the last matching bit, the trie-based algorithms have high IPv4 lookup speed, but limited IPv6 lookup speed for longer prefix lengths. Furthermore, to achieve higher lookup speed, SAIL \cite{Sail} and Poptrie \cite{Poptrie} apply the leaf push method \cite{Expansion} to lookup multiple bits in one memory access. However, they also sacrifice the update speed and the memory cost because of the rule duplications.

\textbf{The binary-based algorithms:}
The binary-based algotihm Hi-BST \cite{Hi-BST} applies Treap to achieve high update speed and low memory cost, but limited lookup speed with too many memory accesses. Furthermore, its fixed scheme to split segments of prefix lengths is only designed for backbone IPv6 rulesets, the edge IPv6 rulesets requires the dynamic scheme to find suitable segments.





\section{Multilayer Online Balanced Tree}

To reduce the memory accesses of IPv6 lookup in the edge network, we propose a multilayer online balanced tree structure MobaTree. In each layer, it constructs a balanced tree to classify non-intersect rules. If some rules are contained by another rule, it further constructs a balanced tree for them in the next layer of this rule. With the multilayer structure, MobaTree performs the binary search with fewer memory accesses.

\subsection{The Online Balanced Tree}

We construct the balanced tree to classify non-intersect rules with different prefixes. When the ruleset contains a few rules, the binary search has fewer memory accesses than traversing bits like SAIL and Poptrie. Without the leaf push method, it also performs faster updates and consumes less memory cost than them. 

\begin{table}[t]\footnotesize
\caption{The ruleset}
\label{rule_set}
\begin{center}
\begin{tabular}{cccc}
\hline
Rule & IP Prefix & Address/Mask & Range  \\\hline
A    & 00*       & 0/2          & 0-63    \\
B    & 01*       & 64/2         & 64-127  \\
C    & 100*      & 128/3        & 128-159 \\
D    & 1011*     & 176/4        & 176-191 \\
E    & 11000*    & 192/5        & 192-199 \\
F    & 111000*   & 224/6        & 224-227 \\\hline
G    & 100011*   & 140/6        & 140-143 \\
H    & 101111*   & 188/6        & 188-191 \\
I    & 001100*   & 48/6         & 48-51   \\
J    & 000011*   & 12/6         & 12-15   \\
K    & 001111*   & 60/6         & 60-63   \\
L    & 0011000*  & 48/7         & 48-49   \\
\hline
\end{tabular}
\end{center}
\end{table}

\begin{figure}[t]
\centering
\includegraphics[width=0.5\linewidth]{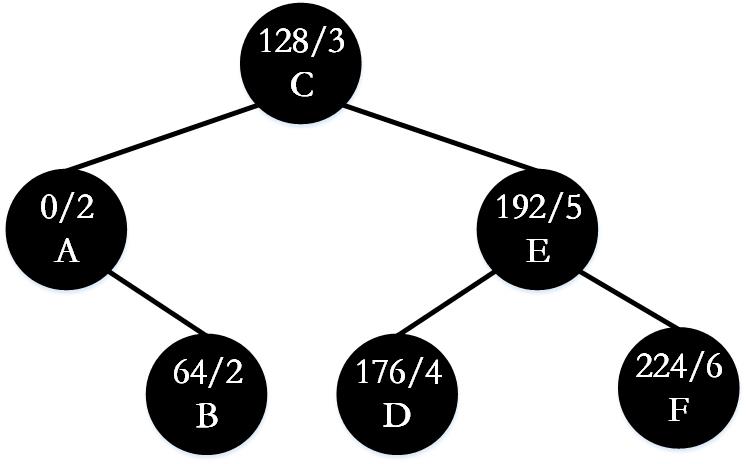}
\caption{The single layer structure of MobaTree.}
\label{MBT_single}
\vspace{-5mm}
\end{figure}

Fig.~\ref{MBT_single} is a balanced tree that stores the forefront six non-intersect rules in Table.~\ref{rule_set}. Assuming the length of IP address is 8 bits, each rule in Table.~\ref{rule_set} can be expressed in the form of IP prefix, address/mask, or range. If the range of each rule does not overlap with other rules, we consider them as non-intersect rules like Rule A-F. To store these rules and perform the binary search, the balanced tree has the following three principles.

\textbf{(1) Each node in the balanced tree stores one rule.} It also represents the IP address range of this rule.

\textbf{(2) Each node contains up to one left child node and one right child node. The child node can recursively contain its child nodes and form a tree structure. The node ranges in the left tree are smaller than the range of this node and that in the right tree are larger than it.} Therefore, a balanced tree can only store rules with non-intersect ranges and perform binary search.

\textbf{(3) The depth of the deepest leaf node is at most twice than the depth of another leaf node.} When we insert or delete a rule in a balanced tree, it will recursively rotate the nodes to maintain this balanced characteristic.

With the balanced tree structure to store $n$ non-intersect rules, the complexity of lookup is only $O(log_2(n))$. It also has the same complexity to support frequent online updates. Furthermore, it consumes low memory cost without rule duplications.

\subsection{The Multilayer Online Balanced Tree}

If some rules are contained by another rule, we will further construct a balanced tree in the next layer of this rule. As shown in Fig.~\ref{MBT_single}, because the ranges of non-intersect rules are sortable, we can construct a balanced tree to store them. However, some rules in Table.~\ref{rule_set} overlap with each other. \textbf{For the IP prefix ranges, if two rules overlap, one must contain the other rather than partial overlap.} Therefore, we can incursively build the balanced tree in the next layer with the following principle.

\textbf{(4) If some rules are contained by another rule, they will be stored in the next layer of this rule and forms a multilayer balanced tree.} If there are still overlaps among these rules, we will incursively build the next layers.

\begin{figure}[t]
\centering
\includegraphics[width=0.5\linewidth]{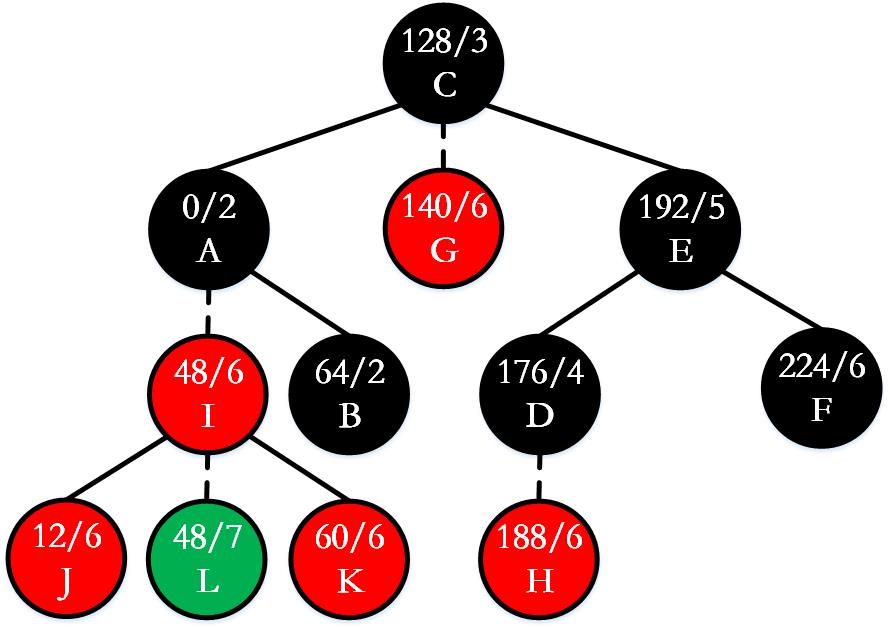}
\caption{The multilayer structure of MobaTree.}
\label{MBT_multiple}
\vspace{-5mm}
\end{figure}

Fig.~\ref{MBT_multiple} shows the multilayer balanced tree that stores all twelve rules in Table.~\ref{rule_set}. (1) Because the forefornt six rules A-F do not overlap with each other and are not contained by another rules, they can be stored in the first layer of MobaTree. The colors of corresponding tree nodes are black in the figure. (2) Because G is contained by C, H is contained by D, and I-K are contained by A, these rules are stored in the next layers of corresponding rules. Each next layer also contains a balanced tree, such as nodes I-K. The colors of these tree nodes in the second layer are red in the figure. (3) Even though Rule L is contained by Rule A, it also contained by Rule I with longer prefix length. Therefore, Rule L should be stored in the next layer of Rule I rather than Rule A. The color of this tree node in the third layer is green in the figure.

\subsection{The Lookup Process}

With the multilayer balanced tree structure, the lookup process contain two steps. First, in each layer of MabaTree, the packet incursively traverses to the left or right child node until it finds the matching node. Second, if the matching node contains the next layer, the packet attempts to further find a matching rule with longer prefix length. Otherwise, we return the rule in this matching node as the matching rule with the longest prefix length. 

The pseudocode for the lookup process of MobaTree is shown in Algorithm.~\ref{node_lookup}. Lines 1-2 initialize the matching rule and the node to traverse. Lines 3-16 incursively find the matching rule with the longest prefix length. If the IP of packet is smaller than the range of this node, we traverse to the left child node in lines 4-5. Similarly, if that is larger than the range of this node, we traverse to the right child node in lines 6-7. Otherwise, we find a matching rule in this node because the IP of packet is in this range. If this node does not contain the next layer, we stop the lookup process in lines 10-11. Otherwise, we attempt to find the matching rule with longer prefix length in lines 12-13. Finally, the lookup process always returns the matching rule with the longest prefix length in line 17.

\subsection{Compared to Treap}

\begin{figure}[t]
\centering
\includegraphics[width=0.45\linewidth]{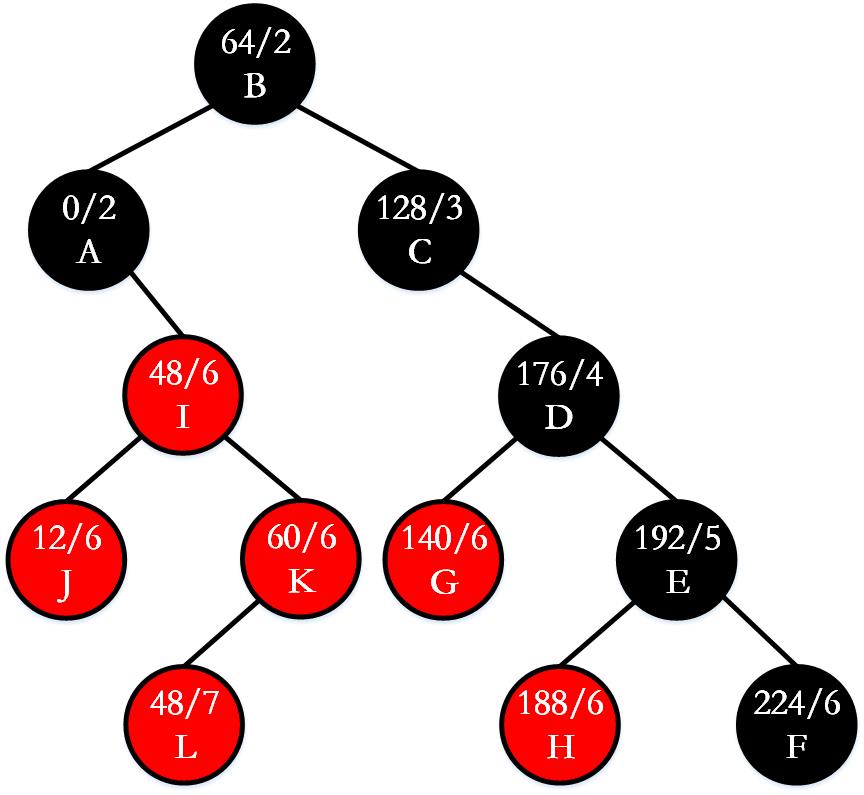}
\caption{The tree structure of Treap.}
\label{Treap}
\vspace{-2mm}
\end{figure}

Compared to the balanced tree in Hi-BST \cite{Hi-BST}, MobaTree has more advantages to support high-speed IPv6 lookup. As shown in Fig.~\ref{Treap}, Hi-BST implements Treap structure to stores rules with different prefix lengths. The value of each node is the begin IP address of range, the values in the left child tree are smaller than it and that in the right child tree are equal or larger than it. Furthermore, the prefix length of each node should be equal or smaller than its child nodes. When a packet lookups, MobaTree has the following three advantages than Treap.

(1) If the non-intersect rules have different prefix lengths, Treap will form a long node chain. For example, the forefront six rules A-F are stored in the black nodes in Fig.~\ref{MBT_multiple} and Fig.~\ref{Treap}. Because the prefix length of each node should be equal or smaller than its child nodes in Treap, it will form a long node chain like node B-F and the tree height is 5. However, MobaTree is more balanced with tree height 3.

(2) When lookup among non-intersect rules, MobaTree has less memory accesses than Treap. For example, when a packet with IP address 192 matches Rule E in Treap, it will further traverse the right child tree and attempt to find a matching rule with longer prefix length. However, MobaTree can return Rule E immediately because this node does not contain a next layer.

(3) When lookup among itersect rules, MobaTree still has less memory accesses than Treap. For example, when a packet with IP address 140 matches Rule G, it will first match Rule C in both balanced trees. In Treap, it will further traverse to Rule D and then find the Rule G with longer prefix length. However, MobaTree will directly match Rule G in the next layer of Rule C.

\begin{algorithm}[t]
\caption{MoBaTreeLookup}
\label{node_lookup}
\begin{algorithmic}[1]
\REQUIRE the root of tree $root$, the packet $packet$;
\ENSURE the matching rule $rule$;

\STATE $rule = NULL$;\

\STATE $node = root$;\

\WHILE {$node != NULL $}
    \IF {$packet.ip < node.ip\_begin$}
        \STATE $node = node.left\_child\_node$;
    \ELSIF {$packet.ip > node.ip\_end$}
        \STATE $node = node.right\_child\_node$;
    \ELSE
        \STATE $rule = node.rule$;

        \IF{$node.next\_layer == NULL$}
            \STATE $break$;
        \ELSE
            \STATE $node = node.next\_layer$;
        \ENDIF
    \ENDIF
\ENDWHILE

\RETURN $rule$;

\end{algorithmic}
\end{algorithm}


\section{Dynamic Programming for Splitting Segments}

To further improve the lookup speed, we split the ruleset into a few segments, which apply the hash table to form multiple small trees. Even though MobaTree is much faster than Treap, the binary search scheme still has too many memory accesses for the large tree. Therefore, we should split the large tree into multiple small trees.

\subsection{The Hash Tables of Segments}

\begin{figure}[t]
\centering
\includegraphics[width=0.6\linewidth]{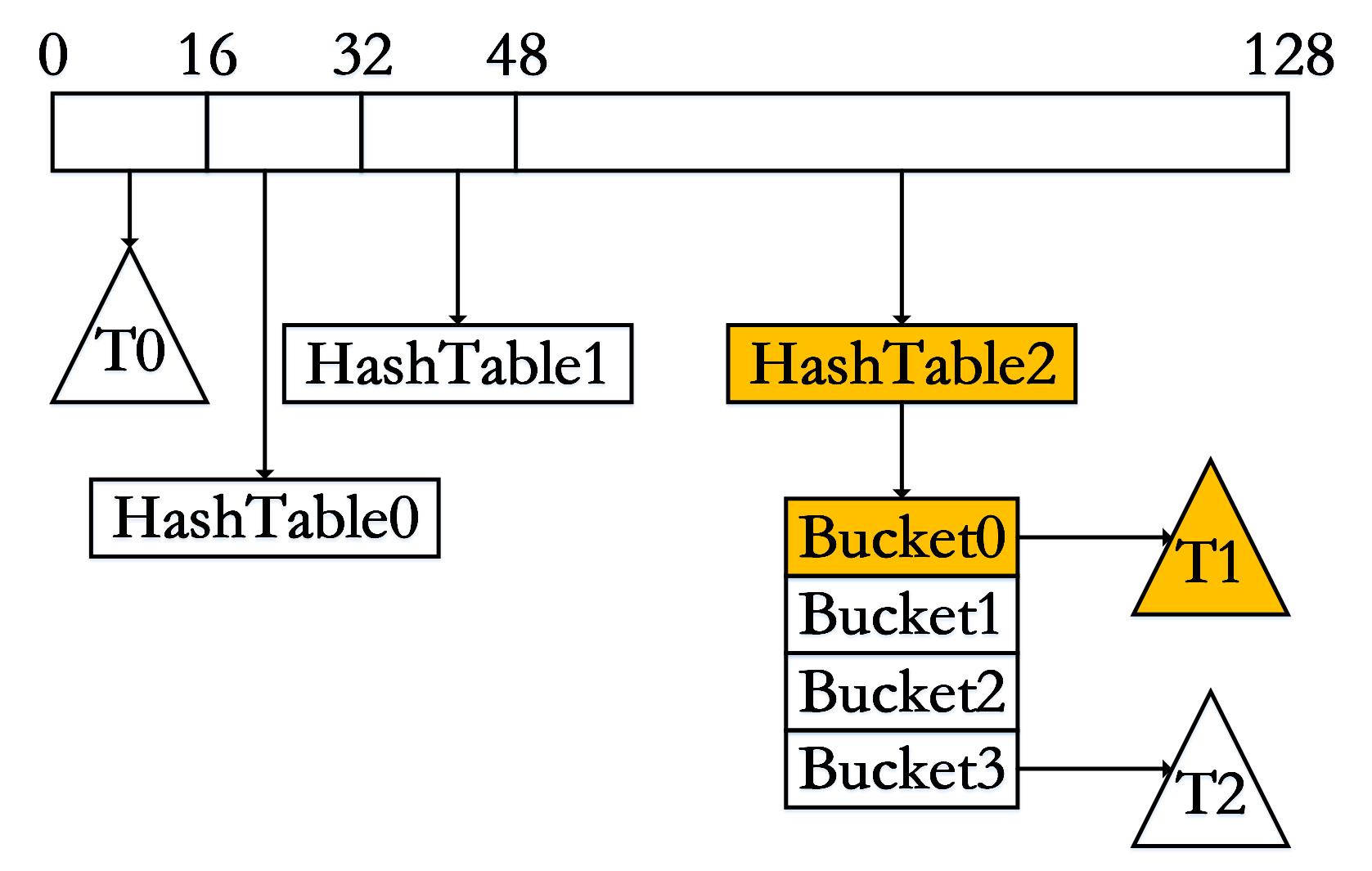}
\caption{The segments to split prefix lengths.}
\label{HashTable}
\vspace{-5mm}
\end{figure}

As shown in Fig.~\ref{HashTable}, we apply the hash table to split the ruleset into segments with multiple small trees. At first, the 128 bits of IPv6 address are split into multiple segments like 0-15, 16-31, 32-47, and 48-128. Each segment contains a hash table to store rules. Then, the prefix of each rule is reduced to the shortest length in its segment. The rule uses its reduced prefix as the key to find the corresponding bucket. Finally, each bucket contains a MobaTree to classify rules with the same reduced prefix. Note that the hash table has hash collision, rules with different reduced prefixes may be indexed to the same bucket. In this case, MobaTree can still distinguish these rules with binary search and achieve high lookup speed.

Because the packet traverses hash tables and lookups in trees, we need to tradeoff between the number of hash tables and the scale of trees to reduce memory accesses. It also means how to split the 128 bits into multiple suitable segments. If the number of segments is too much, the packet requires traversing more hash tables. Otherwise, we forms large trees with more memory accesses. Furthermore, the distribution of rules in the edge network may vary in different environments. Therefore, we will estimate the lookup cost of each segment, then exploit the dynamic programming method to split the 128 bits into multiple suitable segments.

\subsection{The Lookup Cost of Each Segment}

The estimation of lookup cost mainly calculates the memory accesses for the lookup process, and the memory accesses consist of two parts: the hash table and the balanced tree.

\textbf{The memory accesses of hash tables:} When the packet lookups in a segment, it will lookup in the hash table with one memory access. Considering that the packet traverses the segments from long to short, if it matches a rule in a segment, it can skip the lookup process in the remaining segments. In other words, the memory access number of a hash table is equal to the number of rules with shorter prefix lengths. For example, if we split the prefix lengths of Table.~\ref{rule_set} into three segments 0-2, 3-4, and 5-8, we can calculate their memory access numbers of hash tables. If there are 12 packets that match 12 rules respectively, they will all lookup in the segment 5-8. 8 of them can find their matching rules and skip the remaining segments. Therefore, the rest 4 rules with prefix lengths 0-4 will lookup in the segment 3-4. Similarly, only 2 rules will lookup in the segment 0-2. Finally, the memory access numbers of hash tables in these three segments are 2, 4, and 12 respectively. 

\textbf{The memory accesses of balanced trees:} If the matching rule is in this segment, the packet will lookup in a balanced tree. Therefore, we will calculate the sum of lookup cost of trees in this segment. For each MobaTree that contains $n$ rules, the lookup cost is $(\lceil log_2(n) \rceil + 1) * n$, which is the complexity of binary search multiplies the number of rules. For example, segment 2-4 contains four rules A-D and their reduced prefixes are 00*, 01*, 10*, and 10*. Because a balanced tree stores rules with the same reduce prefix, there will be three trees with 1, 1, and 2 rules. Therefore, their lookup costs are 1*1, 1*1, and 2*2 respectively and the sum of lookup cost in this segment is 6. For the packet can not match rules in this segment, it will be indexed to a null bucket or traverse a balanced tree with few memory accesses. Therefore, we ignore the lookup cost in this situation.

\begin{figure}[t]
\centering
\includegraphics[width=0.7\linewidth]{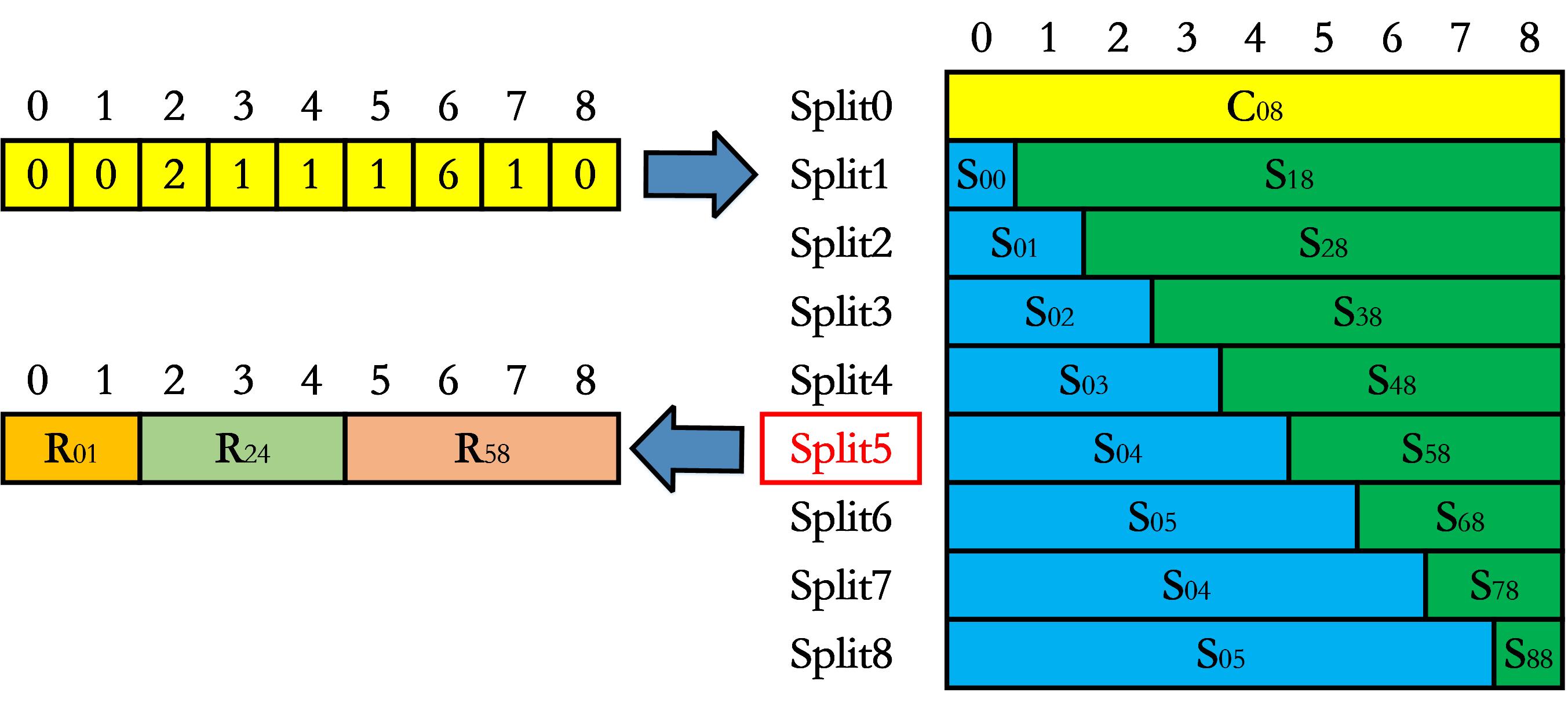}
\caption{The dynamic programming process.}
\label{dynamic_programming}
\vspace{-3mm}
\end{figure}

\begin{algorithm}[t]
\caption{The dynamic programming}
\label{dp}
\begin{algorithmic}[1]
\REQUIRE the lookup cost of each segment $C$;
\ENSURE the minimal sum of lookup costs for segments $S$;

\FOR{$w=0$; $w \le 128$; $w++$}
    \FOR{$x=0$; $x \le 128 - w$; $x++$}
    \STATE $y=x+w$;\
    \STATE $S[x][y]=C_[x][y]$;\
    \FOR{$k=x$; $k < y $; $k++$}
    \STATE $S[x][y]=min(S[x][y],S[x][k]+S[k+1][y])$;\
    \ENDFOR
    \ENDFOR
\ENDFOR
\RETURN $S$;
\end{algorithmic}
\end{algorithm}

\subsection{The Dynammic Programming Method}

To reduce the memory accesses in the lookup process, we should tradeoff between the number of hash tables and the scales of trees. Therefore, we propose the dynamic programming method to split 128 bits into multiple suitable segments. Because the lookup cost of each segment is independent, we just need to find a set of segments with the minimal sum of lookup costs. 

As shown in Fig.~\ref{dynamic_programming}, we apply the dynamic programming to split 8 bis of rules in Table.~\ref{rule_set}. The numbers of rules with different prefix lengths are shown in the left upper coordinate. We use R$_{xy}$ to denote a segment that applies a hash table to store rules with prefix lengths in range $[x,y]$. C$_{xy}$ is the lookup cost of this segment. S$_{xy}$ is the minimal sum of lookup costs for multiple segments, which store rules with prefix lengths in range $[x,y]$. When S$_{xy}$ only contains one segment, it is equal to C$_{xy}$ such as Split0. Otherwise, we attempt to split $[x,y]$ into two parts to minimize the lookup cost such as Split1-Split8. In this case, the lookup cost of S$_{xy}$ is the sum of two parts and the segments of S$_{xy}$ are the integration of them. For example, if we calculate S$_{08}$ and Split5 can achieve the minimal lookup cost, the segments of S$_{08}$ are the integration of S$_{04}$ and S$_{58}$. Through the backtracking to analyze each parts, S$_{04}$ contains R$_{01}$ and R$_{24}$, S$_{58}$ contains R$_{58}$. Finally, R$_{01}$, R$_{24}$, and R$_{58}$ are the suitable segments with the minimal lookup cost to split 8 bits of rules.

Algorithm \ref{dp} shows the pseudocode of dynamic programming. Lines 1-3 enumerate the scale of S$_{xy}$ from small to large. Line 4 indicates that S$_{xy}$ can be C$_{xy}$ itself. Lines 5-7 attempt to split it into two parts to minimize the lookup cost. We return the minimal lookup cost of each S$_{xy}$ in line 10. At last, we can find the suitable segments with the minimal sum of lookup costs through backtracking.

\subsection{Two Optimazation Techniques}

Even though the distribution of rules rarely changes when updating, we also have the scheme to handle this situation. We can use another thread to periodically recalculate the segments and reconstruct the unsuitable segments. Considering that SegMobaTree performs high-speed updates, the lookup process is almost unaffected.

We can also change the weights of rules to suit the distribution of traffic. When calculating the lookup cost of trees and segments, we supposed that rules are matched equally. However, the matching ratios of rules are different in real environments. If the system records the matching ratios of different rules, we can change the weights of rules to calculate the lookup cost. In this situation, rules with high weights will be matched with fewer memory accesses and the lookup performance of algorithm will be further sped up.


\section{Experimental Results}

\begin{figure*}[htbp]
\centering
  \subfigure[The distribution of different rulesets.]{
    \label{rules_distribution}
    \includegraphics[width=0.3\linewidth]{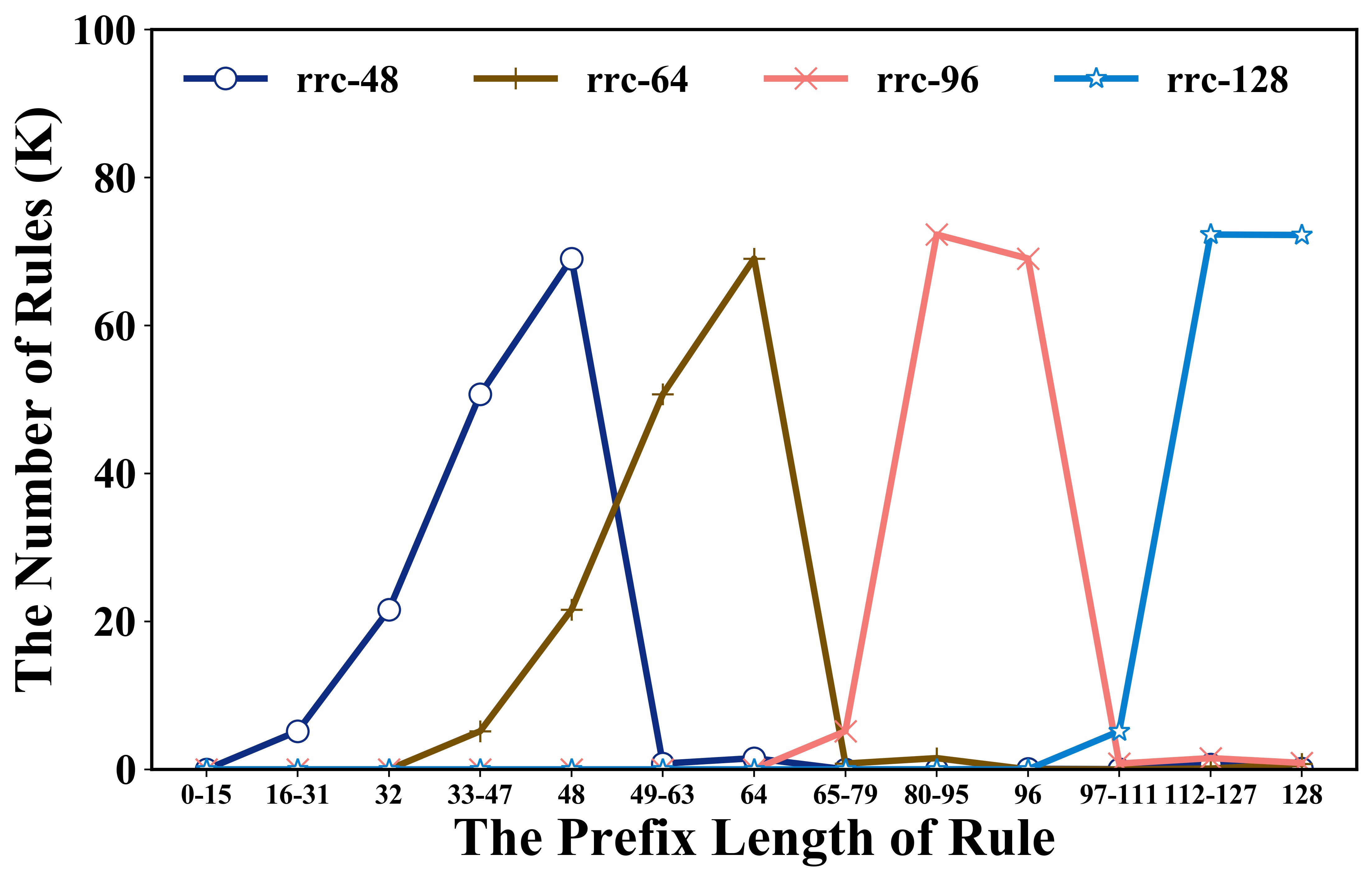}
  }
  \hfill
  \subfigure[The number of memory accesses.]{
    \label{memory_access_num}
    \includegraphics[width=0.3\linewidth]{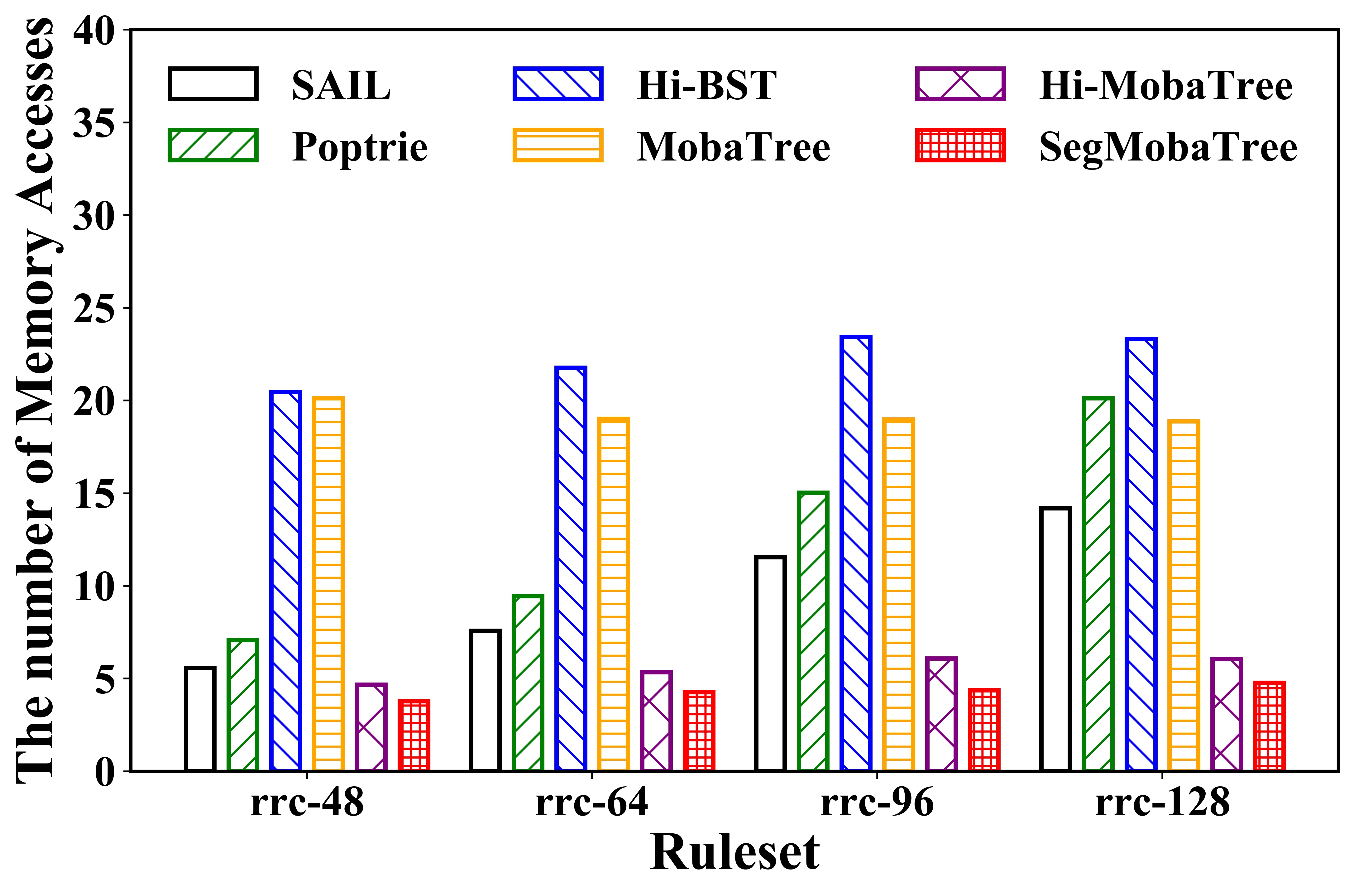}
  }
  \hfill
  \subfigure[The update speed.]{
    \label{update_throughput}
    \includegraphics[width=0.3\linewidth]{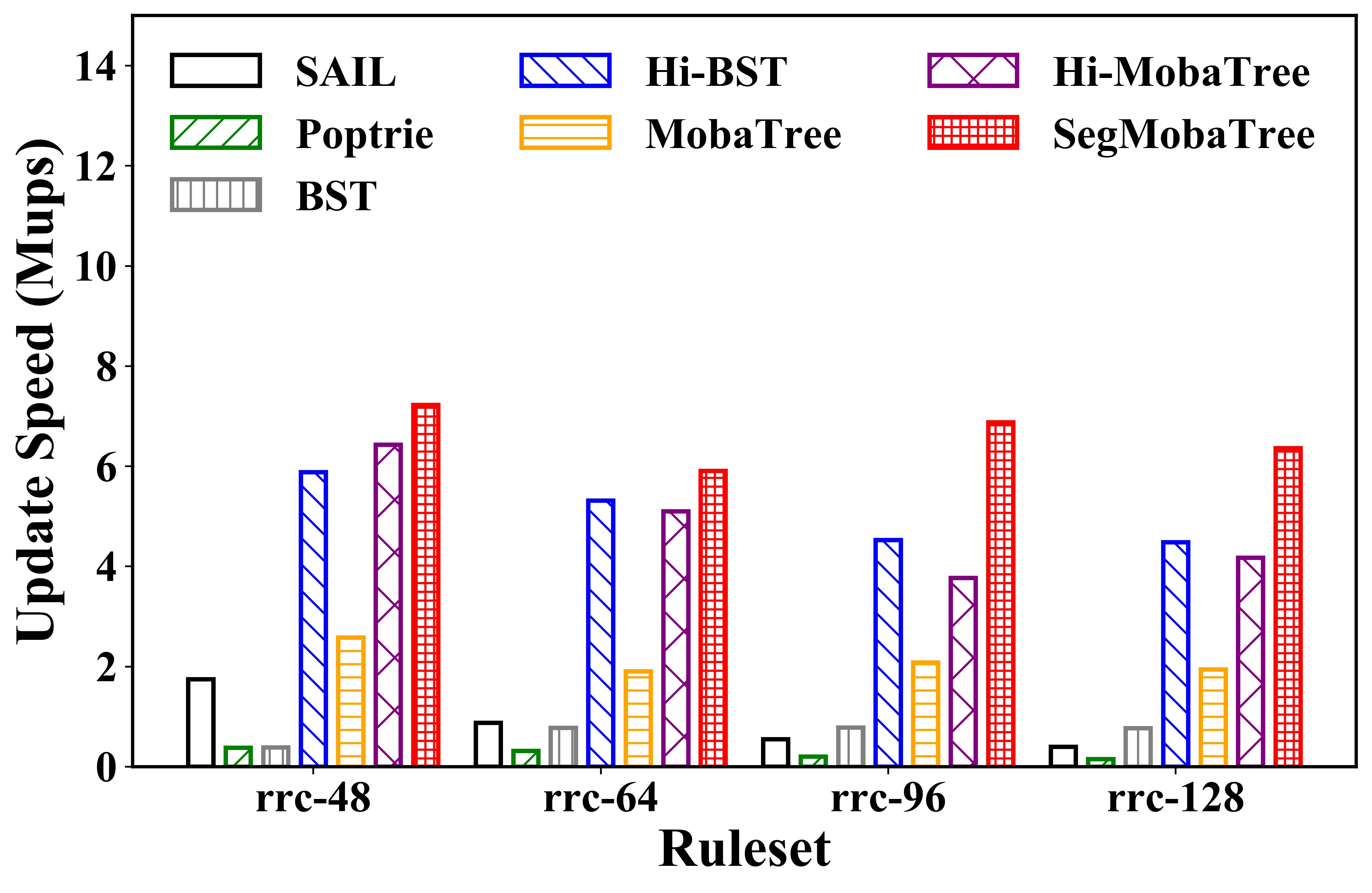}
  }
  \hfill

  \subfigure[The lookup speed for first lookup.]{
    \label{lookup_throughput_first}
    \includegraphics[width=0.3\linewidth]{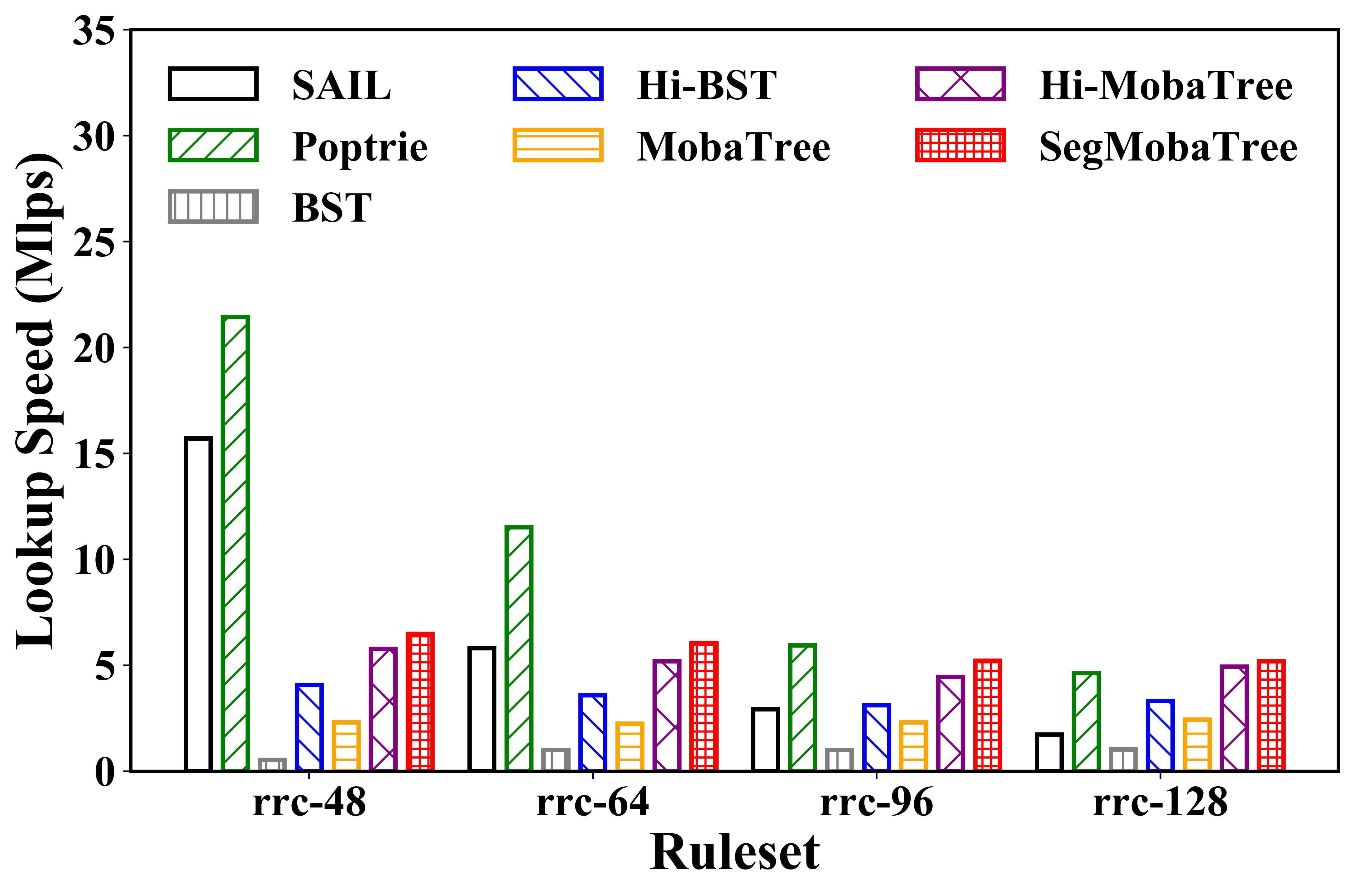}
  }
  \hfill
  \subfigure[The lookup speed for repeated lookup.]{
    \label{lookup_throughput_repeat}
    \includegraphics[width=0.3\linewidth]{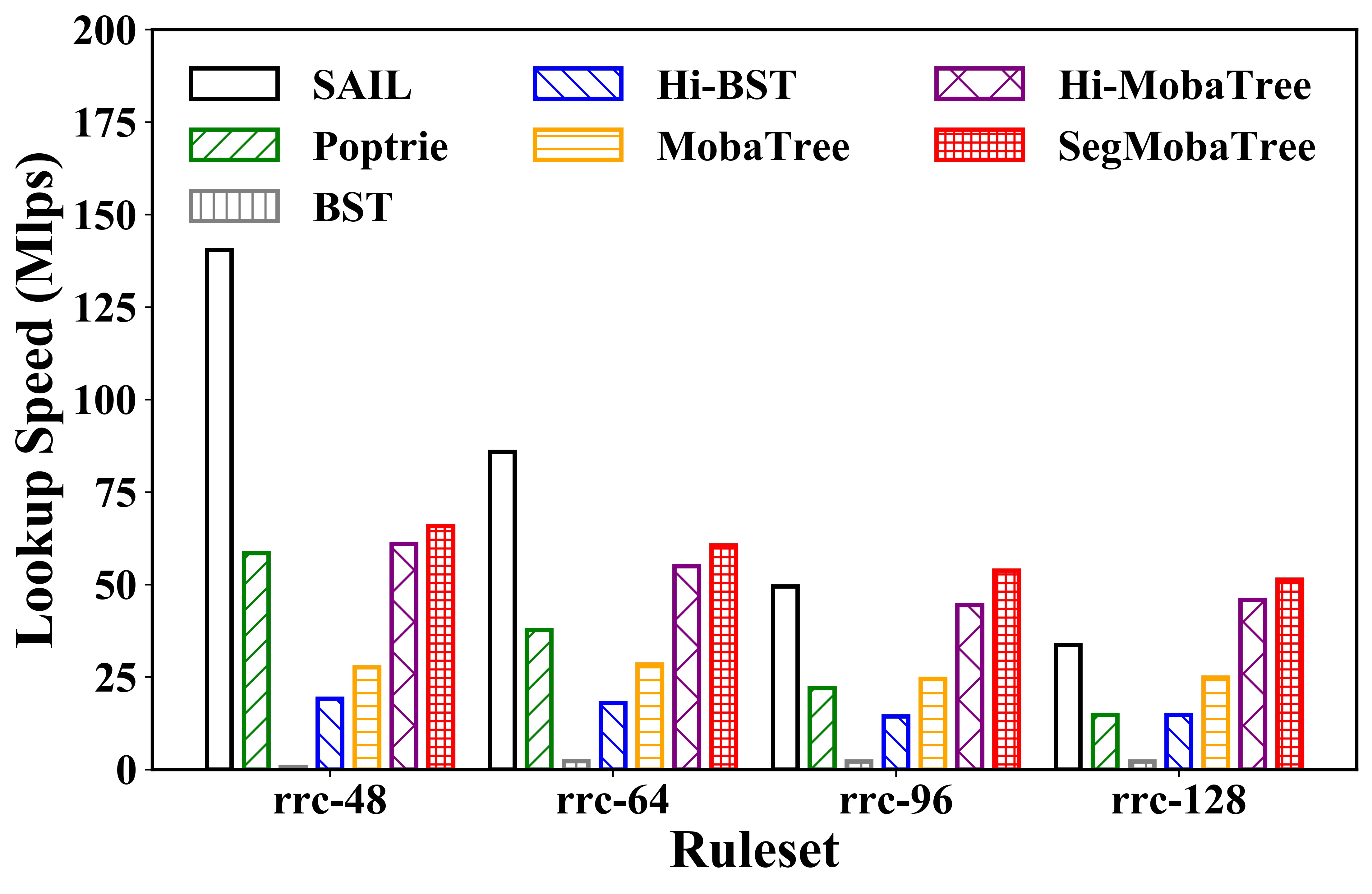}
  }
  \hfill
  \subfigure[The memory cost.]{
    \label{memory_cost}
    \includegraphics[width=0.3\linewidth]{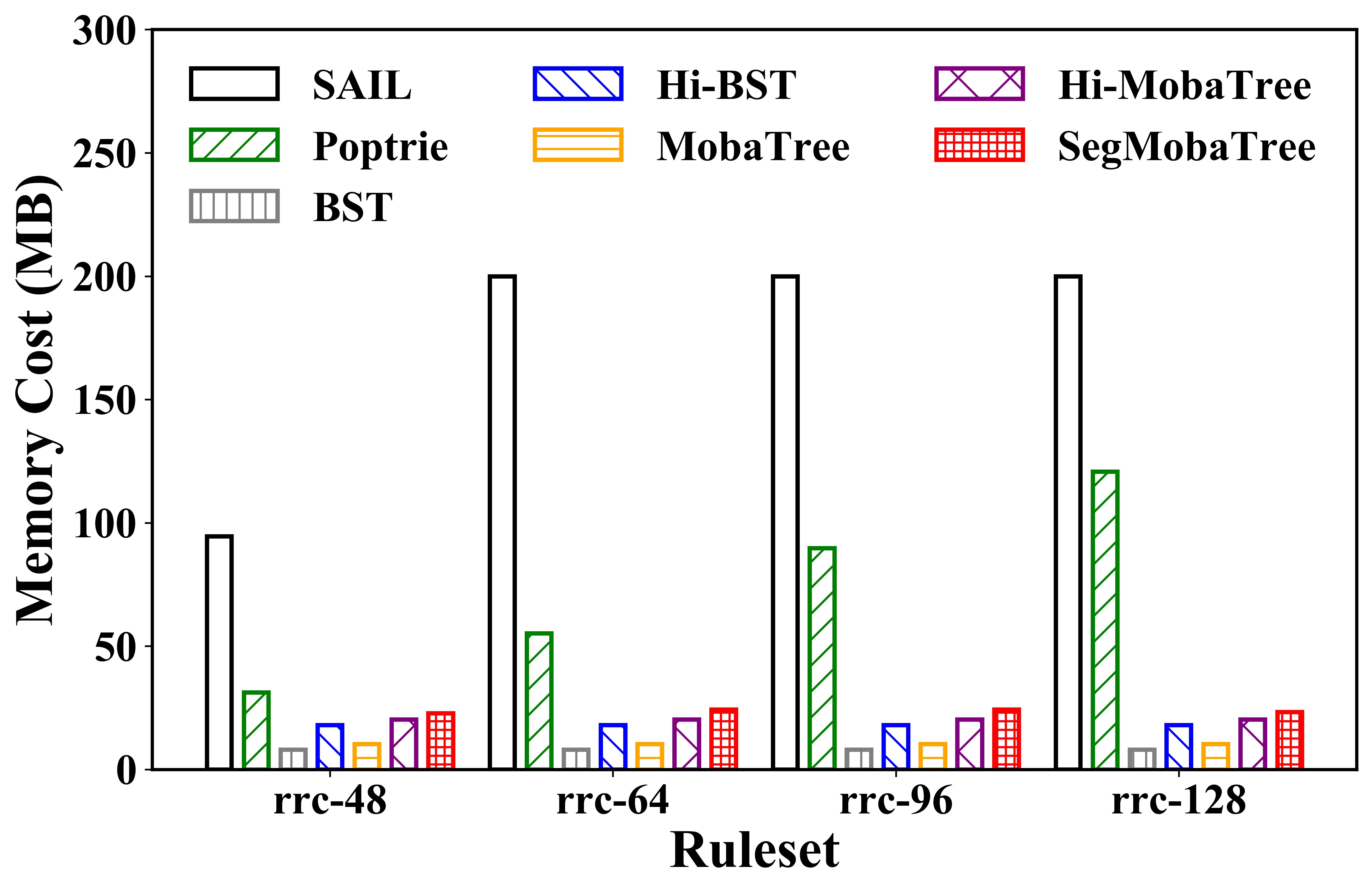}
  }
  \hfill
\caption{The performance of algorithms.}
\label{fig:tuple_performance}
\vspace{-3mm}
\end{figure*}

We compare our SegMobaTree\footnote{https://github.com/zcy-ict/SegMobaTree} with SAIL\footnote{https://github.com/defaning/sail\_algorithm} \cite{Sail}, Poprie\footnote{https://github.com/pixos/poptrie} \cite{Poptrie}, and Hi-BST \cite{Hi-BST}. We implement Hi-BST and modify its fixed segments to adapt the edge IPv6 rulesets. Furthermore, we compare it with BST, MobaTree, and Hi-MobaTree, which is the combination of fixed segments and MobaTree, to show the performance of each component.

Experiments are carried on a computer with Intel(R) Core(TM) i7-8700 CPU@3.20GHz, 6 cores, 64KB L1, 256KB L2, 12MB L3 cache respectively, and 8GB DRAM. The operating system is Ubuntu 16.04.

\subsection{Rulesets and Packet Sets}
We collect the 150K backbone IPv6 ruleset from RIPE \cite{DATA-RIP} and generate the edge IPv6 rulesets according to it. As shown in Fig.~\ref{rules_distribution}, the prefix lengths of backbone IPv6 ruleset rrc-48 mainly focus on 48. However, the prefix lengths of edge IPv6 rulesets can focus on 64, 96, or 128, etc. Therefore, we extend the prefix lengths to simulate edge IPv6 rulesets in different environments. The packet sets are generated according to the rulesets. We use the unordered packet set to measure the lookup speed of the first packet of a flow, then copy each packet 100 times to measure the repeated lookup speed for the flow with the same destination IP address.

\subsection{The Performance of Algorithms}

\textbf{The number of memory accesses:}
Fig.~\ref{memory_access_num} shows the average number of memory accesses. SAIL and Poptrie have more memory accesses when the prefix lengths of edge rulesets are longer. Because BST has hundreds of memory accesses, we ignore it in the figure. With the efficient binary search scheme, the memory access number of MobaTree is even less than Hi-BST. Through the dynamic programming to split segments rather than fixed segments, SegMobaTree has less memory access than Hi-MobaTree. Compared to SAIL, Poptrie, and Hi-BST, SegMobaTree reduces 46.0\%, 57.4\%, and 80.4\% memory accesses in edge IPv6 rulesets.

\textbf{The memory accesses number in the worst case:}
Table.~\ref{memory_access_worst} shows the memory accesses number in the worst case. Because the trie-based algorithms SAIL and Poptrie check multiple bits in one memory access, they have fewer memory accesses in the worst case. Although the binary-based algorithms have disadvantages in this dimension, we still attempt to improve it. Compared to BST, MobaTree reduces the worst number from 4460 to 44. By implementing dynamic programming to split segments, SegMobaTree further reduces it to 22.

\textbf{The lookup speed:}
The first lookup speeds are shown in Fig.~\ref{lookup_throughput_first} and the repeated lookup speeds are shown in Fig.~\ref{lookup_throughput_repeat}. The lookup speeds of SAIL and Poptrie are high in rrc-48 and rrc-64, but slow in rrc-96 and rrc-128 with longer prefix lengths. SAIL mainly performs well in the repeated lookup and Poptrie mainly performs well in the first lookup. For binary-based algorithms, MobaTree is much faster than BST. By splitting segments to avoid one large tree, SegMobaTree can achieve higher lookup speed. Considering both first and repeated lookup situations, SegMobaTree achieves 1.5x, 1.7x, and 2.6x lookup speed than SAIL, Poptrie, and Hi-BST in edge IPv6 rulesets.


\textbf{The update speed:}
The update speeds are shown in Fig.~\ref{update_throughput}. Because SAIL and Poptrie apply leaf push to improve the lookup speed, their update speeds are slower than the binary-based algorithms. With the efficient multilayer structure, the update speed of MobaTree is much faster than BST. By splitting segments to avoid one large tree, SegMobaTree achieve higher update speed. Compared to SAIL, Poptrie, and Hi-BST, SegMobaTree achieves 11.8x, 32.8x, and 1.3x update speed to support online rule updates.

\textbf{The memory cost:}
Fig.~\ref{memory_cost} shows the memory cost of each algorithm. SAIL and Poptrie consume large memory in edge rulesets and we only draw 200MB in the figure. Even though SegMobaTree has high lookup and update speeds, it also achieves low memory cost as Hi-BST.

\begin{table}[t]\footnotesize
\caption{The memory access number in the worst case}
\label{memory_access_worst}
\begin{center}
\begin{tabular}{ccc}
\hline
\tabincell{c}{\textbf{Algorithm}} & \tabincell{c}{\textbf{The memory access number} \\ \textbf{in the worst case}} \\\hline
SAIL & 15 \\
Poptrie & 21 \\
BST & 4460 \\
Hi-BST & 1053 \\
MobaTree & 44 \\
Hi-MobaTree & 28 \\
SegMobaTree & 22 \\
\hline
\end{tabular}
\end{center}
\vspace{-3mm}
\end{table}


\section{Conclusion}
We propose two key techniques to achieve high lookup and update speeds with low memory cost for IPv6 lookup in the edge network. First, we apply MobaTree to perform the binary search and update with efficient multilayer online balanced tree structure. MobaTree constructs a balanced tree to classify non-intersect rules in each layer, then constructs the next layer for rules contained by another rule. Second, we split rules into multiple segments to replace one large tree with multiple small trees, then apply dynamic programming to find suitable segments, which tradeoff between the number of segments and the scale of trees. The experimental results show that, compared to SAIL, Poptrie, and Hi-BST, SegMobaTree achieves 1.5x, 1.7x, 1.6x lookup speed and 11.8x, 32.8x, 1.3x update speed with low memory cost as Hi-BST.

\bibliographystyle{IEEEtran}
\bibliography{7-reference}

\end{document}